\documentclass[aps,pre,twocolumn,showpacs]{revtex4-1}

\usepackage{graphicx}
\usepackage{latexsym}
\usepackage{amsmath,amscd}
\usepackage{times}

\begin{document}

\title{\bf\noindent On the number of common sites visited by $N$ random walkers}
\author{Lo\"\i c Turban}
\affiliation{Universit\'e de Lorraine, Institut Jean Lamour, UMR 7198, Vand\oe uvre l\`es Nancy, F-54506, France}
%


\begin{abstract}
Majumdar and Tamm [Phys. Rev. E {\bf 86} 021135 (2012)] recently obtained analytical expressions for the mean number of {\em common} sites $W_N(t)$ visited up to time $t$ by $N$ independent random walkers starting from the origin of a $d$-dimensional lattice. In this short note I show how the different regimes and the corresponding asymptotic power laws can be retrieved using the notion of fractal intersection.

\end{abstract}

\maketitle

In their recent work~\cite{majumdar2012}, Majumder and Tamm computed analytically the mean number 
of {\em common} sites $W_N(t)$ visited at time $t$ by $N$ independent random walkers starting from the 
origin of a $d$-dimensional lattice at $t=0$. Three distinct regimes were obtained for the large-$t$
behavior:
\begin{eqnarray}
W_N(t) & \sim & t^{d/2}\quad {\rm for}\,\, d<2 \nonumber\\
&\sim & t^{N-d(N-1)/2} \quad {\rm for}\,\, 2<d< d_c(N) \nonumber \\
&\sim & {\rm const.} \quad {\rm for}\,\, d>d_c(N)=\frac{2N}{N-1}\, .
\label{wnt}
\end{eqnarray}
The exponent governing the asymptotic time-dependence of $W_N(t)$ is continuously varying with $N$ and $d$ between the lower critical dimension $d_c'=2$ and the upper critical dimension $d_c(N)$.
Logarithmic corrections appear exactly at the critical dimensions with $W_N(t)\sim t/[\ln t]^N$ 
in $d=d_c'=2$ and $W_N(t)\sim\ln t $ in $d=d_c(N)$ (with $N>1$).

In this note I show how the long-time power-law behavior of 
$W_N(t)$ given in Eq.~(\ref{wnt}) can be simply recovered using an heuristic argument 
based on the notion of fractal intersection. 

In the problem at hand the common sites visited by the $N$ random walkers belong to a fractal object which is the intersection of the $N$ independent random walks.
Let $d_f$=2 denote the fractal dimension of a single random walk and $\overline{d_f}=d-d_f\geq0$ its fractal codimension.
According to Mandlebrot~\cite{mandelbrot1982} the codimension of the intersection of $N$ independent random fractals is given by
\begin{equation}
\overline{d_f(N)}=\min \left(\,d,N\,\overline{d_f}\,\right)\geq0\,.
\label{codim}
\end{equation}
Below the lower critical dimension $d_c'=d_f$, $\overline{d_f}$ and $\overline{d_f(N)}$ vanish, giving 
\begin{equation}
d_f(N)=d-\overline{d_f(N)}=d\,,\qquad d<d_c'=d_f\,.
\label{ldfn}
\end{equation}
The upper critical dimension $d_c(N)$ corresponds to the equality of the two terms on the right in Eq.~(\ref{codim}) so that
\begin{equation}
d_c(N)=\frac{Nd_f}{N-1}\,.
\label{dcn}
\end{equation}
Above this value $\overline{d_f(N)}=d$, leading to
\begin{equation}
d_f(N)=0\,,\qquad d>d_c(N)\,.
\label{udfn}
\end{equation}
Between the two critical dimensions $\overline{d_f(N)}=N\,\overline{d_f}$ applies and one finds
\begin{equation}
d_f(N)=Nd_f-d(N-1)\,,\qquad d_f<d<d_c(N)\,.
\label{dfn}
\end{equation}
\vglue.5mm 
The radius of a walk, $R(t)$, typically grows as $t^{1/d_f}$ such that
\begin{equation}
W_N(t)\sim R(t)^{d_f(N)}\sim t^{d_f(N)/d_f}\,.
\label{wnrt}
\end{equation}
Collecting these results, one finally obtains
\begin{eqnarray}
W_N(t) & \sim & t^{d/d_f}\quad {\rm for}\,\, d<d_f \nonumber\\
&\sim & t^{N-(N-1)d/d_f} \quad {\rm for}\,\, d_f<d< d_c(N) \nonumber \\
&\sim & {\rm const.} \quad {\rm for}\,\, d>d_c(N)=\frac{d_fN}{N-1}\,,
\label{wntdf}
\end{eqnarray}
in complete agreement with Eq.~(\ref{wnt}) for the random walks with $d_f=2$. 

Note that the logarithmic growth of $W_N(t)$ at the upper critical dimension $d_c(N)$ can be obtained  by working with the fractal density of the intersection \cite{turban2013}
\begin{equation}
\rho_N(r)\sim\frac{d [r^{d_f(N)}]}{d [r^d]}\sim r^{-\overline{d_f(N)}}\,,\qquad r\leq R(t)\,.
\label{rho}
\end{equation}
Then:
\begin{equation}
W_N(t)\sim\int_a^{R(t)}\rho_N(r)r^{d-1}dr\sim\int_a^{R(t)}r^{d_f(N)-1}dr\,.
\label{wntrho}
\end{equation}
At the upper critical dimension $d_f(N)$ vanishes so that:
\begin{equation}
W_N(t)\sim\ln\frac{R(t)}{a}\sim\ln t\,.
\label{wntln}
\end{equation}

These results are expected to apply as well in the case of subdiffusive or superdiffusive diffusion processes~\cite{anomalous} with the appropriate value for the fractal dimension of the $N$ walks, $d_f\neq 2$.
For directed walks, an extension of the rules of fractal intersection to {\em anisotropic} fractals is needed.

\end{document}